\documentclass[12pt]{article}
\usepackage{cases}
\usepackage{amssymb}
\usepackage{latexsym, graphicx}
\usepackage{epstopdf}
\usepackage[export]{adjustbox}
\usepackage{hyperref}
\usepackage{xcolor} 

\usepackage{fancyhdr}
\fancypagestyle{plain}{%
\fancyhead{} 
}

\newtheorem{proposition}{Proposition}

\newtheorem{observation}{Observation}

\title{Controlling the Hidden Growth of COVID-19}
\author{Xiubin Bruce Wang\footnote{Associate Professor. bwang@tamu.edu, xbwang@gmail.com. 979-845-9901.} and Chaolun Ma\footnote{Ph.D. candidate. clm16@tamu.edu.}\\Zachry Department of Civil and Environmental Engineering\\Texas A\&M University\\Collge Station, TX 77845}
\begin{document}
\maketitle
\begin{abstract}
The COVID-19 pandemic has plagued the world for months. The U.S. has taken measures to counter it. On a daily basis, newly confirmed cases have been reported. In the early days, these numbers showed an increasing trend. Recently, the numbers have been generally flattened out. This report tries to estimate the \emph{hidden} number of currently alive infections in the population by using the confirmed cases. A major result indicates an existing infections estimate at about 10-50 times the daily confirmed new cases, with the stringent social distancing policy tipping to the upper end of this range. It clarifies the relationship between the infection rate and the test rate to put the epidemic under control, which says that the test rate shall keep up at the same pace as infection rate to prevent an outbreak. This relationship is meaningful in the wake of business re-opening in the U.S. and the world. The report also reveals the connections of all the measures taken to the epidemic spread. A stratified sampling method is proposed to add to the current tool kits of epidemic control. Again, this report is a summary of some straight observations and thoughts, not through a thorough study backed with field data. The results appear obvious and suitable for general education to interested policymakers and the public.
\end{abstract}
\section{Introduction}
The outbreak of the  cornonavirus COVID-19 is an unfortunate incident in the 21st century that suddenly plagued numerous countries. It has brought much of the world economy to a halt. Countries quickly motioned to put it under control. It is an anxious, agonizing process checking the newly confirmed cases of infection and death each day for such an extended, long time. People have observed Italy, Spain, and other countries after China that have been inflicted by this pandemic. Now the U.S. became the seemingly world epicenter with a total confirmed infections exceeding one and half million\cite{uscdc}. As people watch on the daily spread of this pandemic, few know when this pandemic will be put to an end or whether it will spiral out of control again, although pundits worldwide have publicized many findings and predicts. What the general public observe each day are newly confirmed cases from day to day as well as the cumulative totals. The trend of the newly confirmed cases each day was increasing in the early days and is now pretty flattened out overall under the prevalent shelter-in-place policy\cite{CRC}. What does this trend mean? We try to answer the question and also examine a condition under which a stable, flat trend is sustainable. Such an examination appears especially meaningful as the U.S. and the world approach to massive business reopening.

Additionally, in a later summary of all measures taken as of now,  this report tries to clarify how they each take effect in the control of this pandemic in connection to the condition we identify here.  This report proposes a sampling method for testing the communities. It does not discuss the feasibility of the proposed measures in terms of its technical and fiscal constraints. This report did not result from an intensive study. It does not contain field data but only reveals some inherent structural relationships.

In an attempt to understand the driving factors of the epidemic spread, our perspective is different from most mainstream literature of epidemiology. The literature in epidemiology such as \cite{social,granular_data,patten} generally simulate the interactive process of  epidemic transmissive behaviors in a closed region by also considering population migration in and out in order to duplicate the transmissive process. The simulation realizes the mechanism expressed in an array of partial differential equations \cite{forecast, SID_model}, in which the transmission rate is affected by the percentage of the infected in a nonlinear manner. Similar to the literature in epidemiology such as \cite{difficult, ten_times}, we try to understand and gauge the hidden world of the infected population, including the latent and active ones that are not detected yet.  However, our method to estimate the undetected number of currently latent and active infections is based on the publicized numbers of the daily confirmed cases without having to resort to the epidemic process simulation. Our result is simpler and easier to understand and use. Note that this note treats the latent and active as in one undetected group with a comprehensive infection rate.  We allow a different aggregate infection rate for each day. Our weakness is that this report takes infection rate and detection rate as given parameters. This report does not study their specific values.
Current literature report that the total infected population could be six to ten times the cumulative total confirmed \cite{ten_times}.
Our finding shows that the total currently latent and active cases can be six to sixty times  the incremental daily confirmed  total in today's situation that the daily confirmations seem on a flat trend. This findings implies the total infections from 1 to about 10 times the total confirmed infections. One may also gauge the growth or declination of the infections in light of a necessary and sufficient condition for epidemic control that we have found in this report.

In the remainder of this report, we will detail our discussion and derivation of the result. First, we describe the problem in a technical term that we believe characterizes the epidemic process well.
\section{Basic Results}
\subsection{The problem and basic results}

Table \ref{table:notation} tabulates the notation in the report.
\begin{table}[ht]
\caption{Notation} 
\centering 
\begin{tabular}{c p{12 cm}} 
\hline\hline 
$r_i$ & Infection rate on day $i$, One infection at the beginning of day $i$ results in $1 + r_{i}$ infections at the start of the day $i+1$\\
$p_i$ & Probability of confirming an infection from  all the infected\\
$P_i$ & Total infected population in day $i$, excluding those previously confirmed\\
$N_i$ & Daily confirmed on day $i$\\
$\Delta_i$ & Theoretical number of cases confirmed new on day $i$\\
\hline 
\end{tabular}
\label{table:notation} 
\end{table}
We first introduce the epidemic problem. 

\textbf{Problem statement} There is a group of infected people on day one. Out of them, $N_1$ is confirmed and quarantined. On day $i$, each infected case that is not confirmed and quarantined,  referred to as \emph{hidden cases}, grows at a daily rate of $r_i$ into more cases. At the end of each day $i$, \emph{hidden} cases  each have a probability $p_i$ being detected and put into quarantine while the undetected cases continue the infection the next day. The growth rate of $r_i$ is the net growth rate, which is the difference between the new infections deducted by the recovered and dead. Day by day over a period of time, newly confirmed cases are reported by the number $N_1$, $N_2$, ..., and $N_n$.The study is to use the daily confirmed numbers of infections to estimate the \emph{hidden} infections and their trend of growth or declination. The goal is to propose measures for the epidemic control.

Clearly, there are a number of infections \emph{hidden} out in the communities undetected.  Note that the \emph{hidden} cases here are referred to as \emph{undocumented} in \cite{ten_times}. The number of \emph{hidden} cases is the primary interest here.  Be aware that the growth rate $r_i$ here is different from the transmission rate in the conventional epidemiology literature \cite{ten_times}, where the transmission rate is the number of infections that a currently active one may incur per day during the short span of active period. We assume that the confirmed cases have been perfectly quarantined and have lost their ability of further infection. One may consider their remnant infectious ability, even if under quarantine, is implicitly incorporated into the infectious capability of the hidden cases by allowing for a slightly higher rate $r_i$.  We simplify the process from the perspective of someone outside epidemiology by assuming each infected case has an equal average growth rate, which may not be accurate.  One may take this growth rate as an aggregate measure of all the \emph{hidden} cases as a whole. Note we include the latent also in the \emph{hidden} cases. For COVID-19, some reports declare that the latent cases may also be infectious. The net infection (or net growth) rate is random. It is reasonable to assume a constant average rate with fluctuations (e.g., random outcomes) between days for a period over which the test means, ability, and policy do not change significantly. 

We assume a detection rate $p_i$ on day $i$ to represent a percentage of the infections that are confirmed and quarantined on the day. The detection rate has to do with many measures, such as contact tracing of newly confirmed cases. The detected cases are the confirmed cases that the public observes in the daily report. Each day, the reported cases are the tip of an iceberg, which public opinions and public policies are generally based upon.

Let's show the growth process of the infected population in the hindsight. Assume the epidemic moves forward according to the numbered days, day 1, 2, ..., n, n+1,.... Again, there are $N_1$ new cases confirmed on day 1. 

\begin{table}[ht]
\caption{COVID-19 Growth with Confirmed Cases} 
\centering 
\begin{tabular}{c c c c} 
\hline\hline 
Day Index & Infected Population & Theoretically Detected New&  Observed New\\ [0.5ex] 
\hline 
1 & $\frac{N_1}{p_1}$ & $N_1$ & $N_1$ \\
2 & $\frac{N_1(1-p_1)(1+r_1)}{p_1}$ &  $\frac{N_1(1-p_1)(1+r_1)p_2}{p_1}$  & $N_2$\\
3 & $\frac{N_1(1-p_1)(1-p_2)(1+r_1)(1+r_2)}{p_1}$ & $\frac{N_1(1-p_1)(1-p_2)(1+r_1)(1+r_2)p_3}{p_1}$ & $N_3$\\
...&...&...&...\\
n & $\frac{N_1}{p_1}\prod_{i=1,2,...n-1}(1-p_i)(1+r_i)$ & $\frac{N_1p_n}{p_1}\prod_{i=1,2,...n-1}(1-p_i)(1+r_i)$ & $N_n$ \\
n+1 & $\frac{N_1}{p_1}\prod_{i=1,2,...n}(1-p_i)(1+r_i)$ & $\frac{N_1p_{n+1}}{p_1}\prod_{i=1,2,...n}(1-p_i)(1+r_i)$ &  $N_{n+1}$\\ [1ex] 
\hline 
\end{tabular}
\label{table:growth} 
\end{table}

Clearly, there holds  $P_n=\frac{N_1}{p_1}\prod_{i=1,2,...n-1}(1-p_i)(1+r_i)$, where $n \geq 2$ as  on Table \ref{table:growth}. On Table \ref{table:growth}, Infected Population is the number of infected people including those detected and quarantined on the day, the latent and the active but undetected, excluding the confirmed/quarantined on prior days.  The Theoretically Detected New shows the analytical relationship of the daily confirmed cases to the total infected population. Observed New is the daily confirmed cases in the public report.

The following result becomes obvious.

\begin{equation}
\frac{P_{n+1}}{P_n}=(1-p_n)(1+r_n). \label{pop}
\end{equation} 
If we use $\Delta_n$ as the Theoretically Detected New on day $n$, clearly one can reach  the following result,

\begin{eqnarray}
\frac{N_{n+1}}{N_{n}} && = \frac{\Delta_{n+1}}{\Delta_{n}} \nonumber\\
&&=\frac{(1-p_n)(1+r_n)p_{n+1}}{p_n}. \label{delta}
\end{eqnarray}

Combining Equations (\ref{pop}) and (\ref{delta}), we get a major result as illustrated in the following equation.
\begin{eqnarray}
\frac{P_{n+1}}{P_n}&& = \frac{N_{n+1}p_n}{N_{n}p_{n+1}}\label{popgrowth} \\
&&\approx \frac{N_{n+1}}{N_n}. \label{der}
\end{eqnarray} 

Again, $p_i$ is an outcome of a random variable at a constant (average) rate during a stable epidemic period. To simplify, one may simply assume that $p_n$ and $p_{n+1}$ just represent their expected value. Within two consecutive days, if there are no dramatic changes to the sampling methods and regulating policies for detecting the infected cases, one may assume $p_n \approx p_{n+1}$, which explains the approximation in Equation (\ref{der}) above. 
\begin{observation} \label {triv}
When the newly confirmed cases stay flat from day $n$ to day $n+1$, the total infected cases on the two consecutive days may be deemed to roughly stay flat.
\end{observation}

Again, the total infected population on day $i$ mentioned in Observation \ref{triv} excludes the detected and quarantined on prior days. Observation \ref{triv} is trivial in the sense that an equal number of newly confirmed cases and an equal probability of detection naturally allude to an equal size of infected populations on two days. The above discussion also implies:
\begin{equation}
(1-p_n)(1+r_n)\approx \frac{N_{n+1}}{N_n}. \label{final}
\end{equation}

Our interest is in estimating the total infected population at the end of a day, the vast majority of whom are the undetected, active and latent infections. Let take a look at a special case, which is roughly true in the U.S. today, where $\frac{N_{n+1}}{N_n}\approx 1.0$. In this case, we have $(1-p_n)(1+r_n)\approx 1.0$, on which much the discussion ensued is based. 
we summarize the basic result here.
\begin{proposition}\label{prop}
The sufficient and necessary condition to keep an epidemic from growing is to satisfy the following condition:
\begin{equation}
(1-p_n)(1+r_n) \leq 1.0.  \label{nsc}
\end{equation}
When (\ref{nsc}) is in equality, the epidemic reaches an equilibrium, meaning the total hidden plus newly confirmed remain a stable population. As a special case, at low infection rates such as during shelter-in-place, the required detection rate of the infected is roughly equal to the infection rate in order to control the total infected population from growth. 
\end{proposition}

One knows, with social distancing and shelter-in-place orders, $(1+r_n)$ remains very low and yet above 1.0 as one may assume safely for the nation. Then one can roughly estimate the detection probability $p_n$ needed. Proposition \ref{prop} is intuitive and obvious. If $(1+r_n)$ is very close to 1.0 from values above 1.0, one may say that $p_n$ is close to zero from values also above zero when a stable trend of daily confirmed cases is observed. If $p_n$ is very close to zero, in a conservative estimation, say $p_n \approx 0.01$, the total hidden infected population would be about 100 times larger than the latest average daily confirmed amount. The closer $(1+r_n)$ is to 1.0, the closer $p_n$ is to zero, and the larger the total infected population compared with the stable daily confirmed cases. Note that in our simple discussion here, it is impossible to separate the effect of $r_n$ with $p_n$ in the growth $(1+r_n)(1-p_n)$. Therefore, we have no certainty about the values of $p_n$ and $r_n$ respectively. Researchers can only conduct additional studies in order to have separate estimates of the two variables.

For public information, there are several scenarios about the guess at the total \emph{hidden} cases, ranging from the most conservative to the most aggressive. Note that these guesses of the infected total on the day (excluding prior confirmed) is under the circumstance of having flattened new cases each day.
\begin{itemize}
\item{\textbf{Conservative}} In this case, we assume a growth rate $r_n$ to be as reasonably large as possible. In a scenario in which each patient infects 2.5 others during an active period of 14 days, the daily infection rate appears to be $\frac{2.5}{14}=0.18$\footnote{This is a rough number: Some claim that latent cases are infectious while others doubt;  some concluded the active period is as long as 14 days \cite{14days} while others say shorter or longer \cite{5days, 11point5days, 12point5days}.}, meaning $r_n=1.18$.
In this case, if the total new confirmation remains relatively stable, implying $(1-p_n)(1+r_n)\approx 1.0$, $p_n\approx  0.1515$, the total active cases of infection on that particular day would be $\frac{N_n}{p_n}\approx 6.6 N_n$. 
\item{\textbf{Moderate}} In the case that the infection rate, under fairly strict social distancing and shelter-in-place practices, has dropped to a third of that under the normal unrestricted social life as above, where $(1+r_n )\approx 1.039$,   $(1-p_n)(1+r_n)\approx 1.0$ reveals $p_n \approx 0.038$, implying $\frac{N_n}{p_n}=26.64N_n$ is the total infected on day $n$.

\item{\textbf{Aggressive}} This scenario is one that practices very stringent social distancing and shelter-in-place orders, let's assume that the infection rate is controlled to a tenth of the rate for the unrestricted case, meaning $(1+r_n) \approx 1. 018$. In this scenario, if the total new confirmations each day remains flat (hypothetically), the equation  $(1-p_n)(1+r_n)\approx 1.0$ alludes to $p_n \approx 0.018$. This means that the total infections in the population on day $n$ is $\frac{N_n}{p_n} \approx 56.5N_n$.
\end{itemize}

The above example scenarios, all assuming the epidemic is under control by having the total newly confirmed cases flat over a period of days, indicate a large number of currently active and latent infections in the population ranging from 6 to 50/60 times of the daily confirmations. Today, the daily confirmation within the U.S. has been in the range of twenty to thirty thousand cases.  Our simple result suggests that the total current infections in the communities, excluding the quarantined, probably would be in the range of 0.5 to 1.0 million. In the case that schools and businesses are reopened in the near future, where people reasonably practice social distancing and follow other published preventive guidelines, a new situation between the conservative and moderate scenarios listed above, if the new infection has flattened out or has reached a new equilibrium, the total hidden infections alive on the day  would reasonably be in a range of 10-20 times the daily confirmed cases. Of course, the daily confirmed cases after reopening is expected to be higher than under shelter-in-place.

\subsection{Needed Detection Rate at the New Equilibrium}
With reopening, the infection rate $r_n$ would be larger than when the shelter-in-place was enforced, probably by a few hundred percentage. The growth rate likely will be $kr$ where $k \geq 2$ if we use $r$ for the growth rate during shelter-in-place. In order to have the total infected population non-increasing to form a new equilibrium, in light of Equation (\ref{der}), the newly confirmed cases shall not increase from day to day based on the average trend, which roughly means $\frac{N_{n+1}}{N_n}\approx 1.0$.  Note that under shelter-in-place, one roughly has $r_i \approx p_i$ on day $i$, which corresponds to an equilibrium condition $(1+r)\times (1-p)\approx 1.0$, where $p$ represents the stable detection rate during shelter-in-place. Under business reopening, $(1+r_n)(1-p_n)=(1+kr)\times(1-mp)$ is the new infection growth rate, where $mp$ is the new detection rate required to control the infected population from growing, and $m \geq 1.0$. If we have $m \approx k$, we would have  
 $(1+r_n)(1-p_n)=(1+kr)\times(1-mp)\approx (1+r)\times (1-p)\leq 1.0$. The above logic assumes $p_n$ and $r_n$ are both small and that $p_nr_n\approx 0$. To summarize, the following result appears to hold by itself.

\begin{proposition}
In a switch from shelter-in-place to business re-opening, to prevent the infected population from growing, if the infection rate becomes $k$ times larger,  the probability of having the infected to be detected shall be about $k$ times larger accordingly.
\end{proposition} \label{new-equill}

\textbf{What does this larger detection probability mean?} With social distancing and shelter-in-place, a low infection rate allows an equilibrium with a low detection rate, which is a balance between the new infections (recovery and death are considered a negative increase) and the ones quarantined. With business re-opening, the infection rate would probably spike, say by $k$ times, if the proportion of the population being checked or sampled (or an equivalent of it) remains unchanged as under shelter-in-place, the infected population would explode. The rate of explosion, in light of the discussion at the beginning of this subsection, would be $(1-p_n)(1+r_n) \approx 1+(k-1)r$. There will be no equilibrium between the newly infected and the newly detected if the detection rate falls behind the infection rate. The practical meaning of this observation may be recapped as follows. We do not have a specific value for the daily infection rate $r_n$ before and after the business re-opening, which needs to be specially studied by professionals.  

\begin{proposition}
With business re-opening, if the chance of detecting an infected is not keeping pace precisely with the infection rate increase, the infected population will keep growing.
\end{proposition}

Worthy of a note is that the detection rate here is the probability of a random infection in the community to be tested based on some mechanism (such as social contact tracing, body temperature check at some key locations, random sampling in the communities, etc.). This report is not in a complete agreement with some media reports such as \cite{hv}, which advocates for an absolute increased total number of tests. We think the sample rate or its equivalent is the factor effective to epidemic control. It's about the percentage of the people who shall be tested and quarantined, not the absolute number. Again, stringent social contact tracing is an effective means to raise the percentage of infections to be quarantined.

\section{Strategies Recapped}
In this section, we briefly summarize the current strategies in dealing with COVID-19.

\subsection{Two Categories of Measures}
The previous section makes it clear that to put the epidemic under control, one must have inequality (\ref{nsc}) satisfied. At the very minimum, (\ref{nsc}) shall be an equality. All the measures so far undertaken by governments and other entities fall into two categories: controlling the infection rate and controlling the detection rate.
\begin{itemize}
\item{\textbf{Containing the infection rate}} The measures in this category include: shelter-in-place, specially social distancing, quarantine policy, all means of disinfection or sanitation, medical treatment, etc.  Immunization as the final solution may be considered as a special means,
\item{\textbf{Improving the detection rate}} This category includes social contact tracing, population sampling,  test kits of improved accuracy, additional means such as quick antibody test, body temperature measurement as a pre-screening means for virus test, etc. 
\end{itemize}

It appears that much more can be done at the front of increasing detection rate. The difficulty is in the fact that to identify an infected resembles a witch hunt due to the randomness or uncertainty. Google and Apple are reported working together to develop social media applications for (voluntary) tracing social contact and for alert of potential danger of infection, which serves  both purposes of reducing infection and increasing detection rates\cite{usatoday, verge}. Contact tracing of an infected person remains a valid means. If contact tracing is effective, it equivalently increases the detection rate and decreases the infection rate simultaneously. Stringent contact tracing is a proven means in other countries such as South Korea. As mentioned shortly above, the public body temperature measurement through automatic machines when the public pass through key locations may prove to be an extremely beneficial means to greatly increase the detection rate of the infected. Here, We propose a sampling method as an additional but not a substitute means to the existing measures.

\subsection{A Sampling Method}
In light of condition (\ref{nsc}), a sample rate that is equal to the infection rate, or other means such as very stringent social contact tracing that achieve an equivalent effect,  would work. We mean here for a random sampling of the public. This random sampling may also be coupled with enhanced contact tracing of the community or social groups of newly confirmed infections. We propose a stratified sampling by prioritizing the high-risk groups. There are statistics in the U.S. to show different infection rates for population strata according to ethnicity, age, or other criteria (\cite{uscdc}\cite{CRC}). If the sampling cost is too high, a low-cost screening process may be first implemented such as a temperature check. Temperature check at major traffic locations such as airports and subway stations is an effective pre-screening means. We propose a sample rate proportionate to the infection percentage of the strata weighted by the varying costs, one of the costs being life-threat. Obviously, the aged groups, especially those with complications, have a more severe consequence from the infection.

To understand the necessity of sampling in public, one might consider such a question: Would social contact tracing alone meet the needs of control without sampling of the population? Vast deployment of body temperature measurement machines is an example of large population sampling. There may be a debate here. But in our view, periodic sampling to a certain extent, preferably on a daily basis, is necessary to an enhanced social contact tracing.

\subsection{Vulnerable and High Cost Groups}
\begin{figure}[h!]
    \centering
    \includegraphics[scale=0.3]{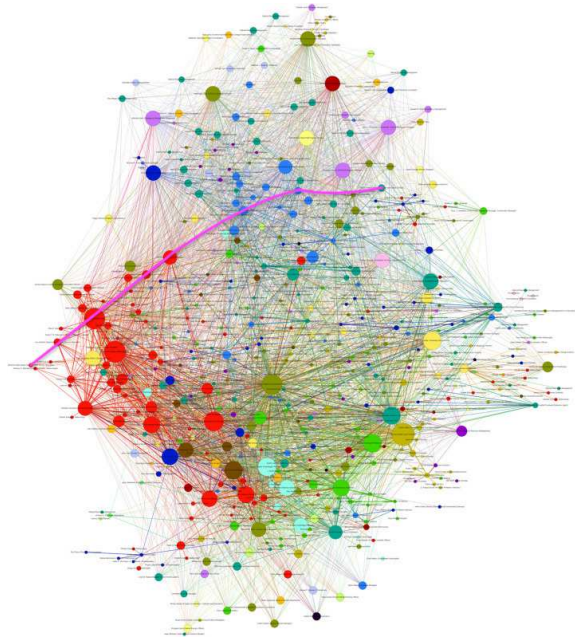}
    \caption{Example Social Contact Network with Entities (Strings) and Nodes (Sources:\cite{sample})} \label{net}
\end{figure}

Consider social networking in the context of COVID-19. Figure \ref{net}  illustrates this general idea. This social network (via proximity connection) contains the most effective entities in terms of spreading the virus. An individual may be an entity shown as a string that connects multiple dots (or nodes). An entity may be a household in which the household members are considered fully communicated internally in terms of virus infection. A string accounting for a household (e.g. the connected links of the network)  may be connected to many other nodes such as shopping malls, schools, plants, etc. There are also nodes that many entities traverse. Nodes and strings all have varying attributes that contribute to the virus spread to different degrees.  Schools, including k-12 and college education when face-to-face classes are resumed may be considered high-risk nodes. Families of active members exposed to close contact with a large number of others are also considered as strings with large weight. The goal of dealing with this network is to identify the nodes and strings that may effectively circumvent the spread of COVID-19.

Although much needs to be studied in future through the academic community regarding COVID-19 spread, much can now be done by governments and people in their responsible positions. These include: identify critical locations with most social connections, identify entities that connect most critical locations (e.g. household with multiple members working at shopping malls, factories, attending schools simultaneously). A major job of the governments and administrations would be to implement measures that impede or even \emph{cut off} the social network to decrease or minimize its epidemic connectivity without sacrificing other major functions of it such as productivity, quality of life, etc.

\section{Conclusion}
This is a preliminary discussion of COVID-19 spread in the process of test and quarantine. The main focus is about the general conditions needed to put the epidemic under control, a situation in which the total infected population size does not grow or even shrink.  The results are intuitive and straightforward.  This document may, therefore, serve as an educational material to the public or the government officials in positions relevant to disease control. 

The major finding is that the effective detection rate should keep up in pace with the infection rate such that $(1+r)(1-p)\leq 1.0$. This finding implies that the detection rate under business re-opening shall keep pace with the expectedly spiking infection rate. This discussion has touched on the means in two categories, detection and infection rates control, respectively. Enhanced social tracing may be an effective means of increasing the detection rate. The detection rate here is the probability of a random infection in a community to be selected and tested for infection Random sampling based on strata is also proposed, which emphasizes the strata likelihood and also strata severity of infection. A principle is that the sample rate shall be consistent with the strata infection rate weighted by infection consequence. Note that direct sampling for infection test may turn out to be too costly. Therefore, vast deployment of automatic body temperature measurement machines at key locations with large volume of traffic, especially traffic that risk transmitting virus among regions or communities is necessary as a pre-screening step of social sampling.

Another interesting finding is that when the daily confirmed cases remain relatively stable a trend, an equilibrium has likely formed between infection and detection. This means $(1+r)(1-p) \approx 1.0$. Under social distancing and shelter-in-place, the infection rate is low, and obviously, the detection rate $p$ is low as well. One may estimate the total alive infections hidden out in the communities of up to 50 times the daily confirmations. Note that the same day total alive infections of 50 times the daily confirmation is equivalent to a total cumulative infections of about 10 times the cumulative daily infections.  This estimate may serve as an alert to the policymakers when they prepare for business re-opening. Those undetected infections would incur a higher infection rate during business re-opening and require a higher detection rate accordingly. The needed higher detection rate would come from a heightened contact tracing system, which is desirably coupled with new community sampling practice.

In fact, the infection rate and detection rate are both random numbers from day to day. Therefore, the ratio between confirmed cases on two consecutive days may fluctuate around 1.0, not stably equal to a constant 1.0. The infection and detection rates may be taken as the expected values and be used for the trend analysis. If a model fully considers the randomness of the two rates, a reliable condition to control the epidemic is likely to be $(1+r)(1-p) < 1.0-\epsilon$, where $\epsilon$ is a small positive number representing a little room to allow for the randomness and errors in the process. Without further extensive studies, the two parameters used here, $r$ and $p$, would not be known exactly. However, the relationship and the qualitative meaning of the results are sound.
This discussion was driven by interest, it did not involve extensive, data rich effort which are often backed by projects. Therefore, the limitations of this report are multitude. We hope this report may bring its audience to an operations research perspective for this epidemic control.

\section*{Acknowledgment}
This study is a self-motivated effort with partial support from a E.B. Snead '25 Career Development Professor I fund through the Zachry Department of Civil and Environmental Engineering at Texas A\&M University.  Errors and mistakes are solely the authors'.


\end{document}